\documentclass[prl,twocolumn,aps,amsmath,amssymb,showpacs,10pt]{revtex4-1}
\usepackage[version=3]{mhchem} 
\usepackage[T1]{fontenc}
\usepackage[utf8]{inputenc}
\usepackage[english]{babel}
\usepackage{units}
\usepackage{amsmath} 
\usepackage{amsfonts} 
\usepackage{amssymb} 
\usepackage{color}
\usepackage{graphicx}
\usepackage{epsfig}
\usepackage{epstopdf}
\usepackage{float}
\usepackage{nicefrac}
\usepackage{url}
\usepackage{amsmath}
\begin{document}

\title{Electron--optical phonon coupling in suspended bilayer graphene}

\author{Antti Laitinen$^{1}$, Manohar Kumar$^1$, Mika Oksanen$^1$, Bernard Pla\c{c}ais$^{2,1}$, Pauli Virtanen$^1$, and Pertti Hakonen$^{1}$}

\affiliation{$^1$Low Temperature Laboratory, O.V. Lounasmaa Laboratory, Aalto University, 00076 AALTO, Finland}
\affiliation{$^2$Laboratoire Pierre Aigrain, Ecole Normale Sup\'{e}rieure, CNRS (UMR8551), Universit\'{e} Pierre et Marie Curie, Universit\'{e} Paris Diderot,  75231 Paris, France}

%
\begin{abstract}

Using electrical transport experiments and shot noise thermometry, we investigate electron-phonon heat transfer rate in a suspended bilayer graphene. Contrary to monolayer graphene with heat flow via three-body supercollision scattering, we find that regular electron - optical phonon scattering in bilayer graphene provides the dominant scattering process at electron energies $ \gtrsim 0.15$~eV. We determine the strength of these intrinsic heat flow processes of bilayer graphene and find good agreement with theoretical estimates when both zone edge and zone center optical phonons are taken into account.
\end{abstract}
\maketitle
%

Electron-phonon coupling has been investigated extensively in monolayer graphene (MLG), both theoretically  \cite{Kubakaddi2009,Tse2009,Bistritzer2009,Viljas2010,Suzuura2002,Katsnelson2012} and empirically in quantum transport experiments \cite{Betz2012,Betz2012a,Yan2012,Laitinen2014}. The weak coupling between acoustic phonons and electrons limits electronic cooling, and and as a result extrinsic processes (``supercollisions'') take over in typical samples \cite{Song2012,Betz2012a}, even in suspended ones  \cite{Laitinen2014}. In supercollisions, the restrictions in energy transfer by single acoustic phonon scattering  \cite{Katsnelson2012} are circumvented via three body collisions, where disorder facilitates for the participation of phonons with a larger momentum in the scattering process. Understanding the scattering processes in monolayer and bilayer graphene is important for designing high-quality graphene transistors, in which the mobility at large bias will be limited by the scattering from optical phonons in the absence of extrinsic processes.

Except for the small energies \cite{McCann2006a}, the electronic band structure of bilayer graphene (BLG) is quite different from the monolayer (MLG): instead of massless Dirac fermions, the bilayer has massive particles as charge carriers. This results to a larger density of states (DOS) of the electrons: the ratio of DOS at energy $E$ is given by $\gamma_1/|E|$, where  $\gamma_1 \approx\unit[0.4]{eV}$ corresponds to hopping between the two layers in bilayer graphene \cite{Misu1979}; the difference can be further amplified by velocity renormalization effects in MLG \cite{Kotov2012}. In addition to increasing the general electron-phonon heat flow in bilayer as compared to monolayer, these differences also turn out to contribute toward increasing the relative importance of scattering from optical phonons in bilayer graphene.

In this work, we have employed shot noise thermometry and conductance measurements to determine the electron-phonon coupling in high-quality, suspended bilayer graphene at large bias voltages. We demonstrate that in bilayers we can reach the intrinsic behavior at high bias, and that the electron-phonon scattering is governed by optical phonons. At bias voltages corresponding to electronic temperatures $> 300$ K, we find strong enhancement of the electron phonon coupling due to optical phonons which results in a typical thermal activation -type of  growth of the heat flow (cf. Fig. 1). The magnitude of the power flow can be well accounted for by the existing theories for a bilayer when long wave length longitudinal (LO)  and transverse (TO) optical  modes around zone center (ZC, $\Gamma$-point) are taken into account with additional contributions from zone edge modes (ZE, K-point). In the regime of optical phonon scattering, only a weak dependence on chemical potential is observed, consistent with theory \cite{Bistritzer2009,Viljas2010}.

\begin{figure}
  \includegraphics[width=0.9\linewidth]{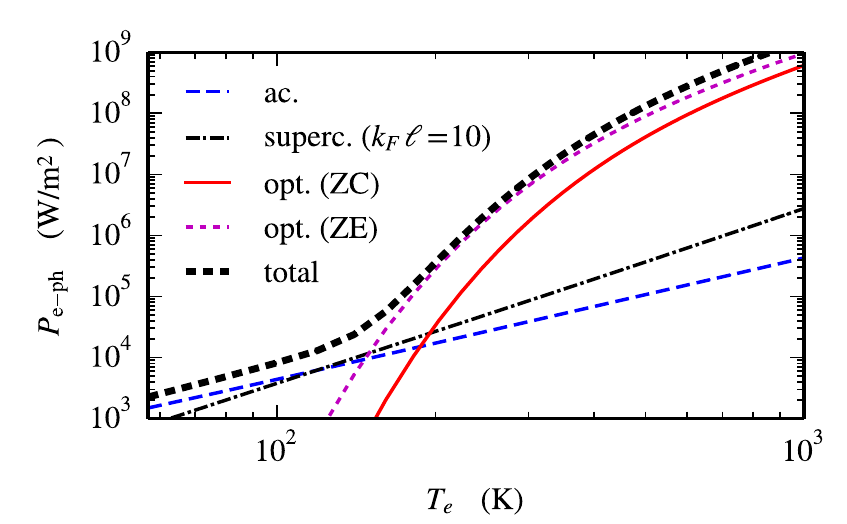}
  \caption{\label{fig:bi-J-comparison}
    Theoretical estimates for the power density from electrons to phonons
    in bilayer graphene as a function of electronic temperature $T_e$
    (for $n=\unit[10^{11}]{cm^{-2}}$ and $T_{ph}=0$), due to different phonon scattering processes.
    Acoustic and optical ZC phonon results are from Eqs.~(21) and (30) in
    Ref.~\onlinecite{Viljas2010}, assuming gauge potential $D_2=\hbar{}v_F\beta/(2a)\approx\unit[7]{eV}$ and
    screened deformation potential \cite{Ochoa2011} $D_1(q=k_F)=\unit[0.6]{eV}$.
    Supercollision and ZE estimates are as from Eqs.~\eqref{eq:ze},\eqref{eq:supercollision}
    below.
  }
\end{figure}

Several different phonon scattering processes are expected to contribute to the transfer of heat from the charge carriers, \textit{ie.} electrons or holes, to the lattice in BLG. First, estimate for the effect of collisions with acoustic phonons can be characterized by a power law  $P=\Sigma (T_e^{\delta} -T_{ph}^{\delta})$, where $T_e$ is the electron temperature, $T_{ph}$ the phonon temperature, $\Sigma$ the coupling constant and $\delta$ a characteristic exponent \cite{Giazotto2006}. The maximum change of momentum at the Fermi level is twice the Fermi momentum $2 k_F$, which corresponds to phonon energy $\hbar \omega_{2k_F}$. This energy defines a characteristic temperature, the Bloch-Gr\"uneisen temperature, $k_B T_{BG}=\hbar \omega_{2k_F}\approx\unit[18]{K}\times\sqrt{n/\unit[10^{11}]{cm^{-2}}}$, above which only a fraction of acoustic phonons are available for scattering with electrons in the thermal window. Our bilayer graphene experiments have been conducted near the Dirac point at charge densities $n < 0.1-3.4 \cdot 10^{11}$ cm$^{-2}$, which corresponds to $T_{BG} < 34$ K for longitudinal acoustic phonons. Using the Kapitza resistance from Ref. \onlinecite{Song2014}, we find that all our high-bias results have been measured in the regime of $T > T_{BG}$, where the scattering of electrons from acoustic phonons leads to $\delta = 1$ or $\delta = 2$, depending whether $k_B T_e << \mu$ or $k_B T_e > \mu$, respectively \cite{Viljas2010}. Here $\mu$ denotes the chemical potential.


The power density to the zone center (ZC) longitudinal and transverse optical modes in BLG, which have energies $\Omega_{ZC}\approx\unit[0.2]{eV}$, can be estimated as~\cite{Viljas2010}
\begin{equation} \label{e.blqlto}
{P}_{e-op}^{(ZC)}
=\frac{18A\Omega_{ZC}^3 (\gamma_0')^2\hbar}{\pi (\hbar v_0)^4\rho}
\frac{\gamma_1}{\Omega_{ZC}}
[n_e(\Omega_{ZC}) - n_{op}(\Omega_{ZC})]\mathcal{G}(\mu,T_e).
\end{equation}
Here, $A $ is the sheet area, $\gamma_0' = 42$ eV/nm
\cite{Yan2007}, $\rho=4M_CN/A$, $v_0\approx\unit[10^6]{m/s}$, and
$n_e(\Omega_{ZC})$ and $n_{op}(\Omega_{ZC})$ are Bose distribution functions
evaluated at temperatures $T_e$ and $T_{ph}$. Finally,
\begin{equation} \label{e.gfu}
\mathcal{G}(\mu,T_e) = \int_{-\infty}^{\infty}\frac{1}{4}(|x|+|x-1|)
[f(\Omega_{ZC}(x-1)) - f(\Omega_{ZC} x) ]dx
\,,
\end{equation}
describes the dependence on the chemical potential. We assume that the coupling between optical and acoustic phonons is not limiting the energy flow \cite{Bonini2007,wu2010}, and consider the optical phonon temperature $T_{ph}$ as a constant.

Intervalley scattering by zone edge (ZE) optical phonons also contributes to the heat current. In MLG, the ZE point optical modes dominate over ZC phonons in resistance \cite{Mauri2014,Fang2011}. The results of Ref.~\cite{Mauri2014} indicate $\sum\langle{M^2_{ZE,j}}\rangle/\sum\langle{M^2_{ZC,j}}\rangle=\Omega_{ZC}/\Omega_{ZE}\approx1.3$ for the ratio of the angle-averaged squared matrix elements, which are relevant for the heat current. The corresponding power density is obtained by substituting $\Omega_{ZC}$ with $\Omega_{ZE}$ in Eqs. \ref{e.blqlto} and \ref{e.gfu}.

The total heat flow by optical phonons is the sum of these two contributions: $P_{e-op} = P_{e-op}^{(ZE)} + P_{e-op}^{(ZC)}$. Unfortunately, we are not aware of microscopic results for electron-ZE-phonon coupling in BLG. We can, however, obtain a rough estimate by assuming that the ratio of matrix elements is similar in BLG as in MLG, in which case the ZE contribution becomes
\begin{align}
  \label{eq:ze}
  P_{e-op}^{(ZE)}\approx{}\frac{\Omega_{ZC}}{\Omega_{ZE}}\times{}P_{e-op}\rvert_{\Omega=\Omega_{ZE}}
  \,.
\end{align}
At temperatures $T=\unit[300\ldots1000]{K}$, the two contributions are of the same order of magnitude, $P_{e-op}^{(ZE)}=4\ldots1.5P_{e-op}^{(ZC)}$.

Finally, the effect of acoustic phonon supercollisions in BLG can be estimated similarly as derived for MLG in Ref.~\onlinecite{Song2012}.  Within the quadratic dispersion approximation and a screened (Thomas-Fermi) BLG electron-phonon interaction model \cite{Ochoa2011}, the power density becomes
\begin{align}
  \label{eq:supercollision}
 {P}_{e-ph}
  &\approx
  \frac{  9.62
    \tilde{g}^2 \nu_1^2
  }{
    2 \hbar (k_F\ell)_{MLG}
  }
  k_B^3
  (T_e^3 - T_{\rm ph}^3)
  \,,
  \;
  \tilde{g}^2
  =
  \frac{
    \frac{D_1^2}{4\alpha^2} + D_2^2
  }{
    2\rho s_L^2
  }
  \,,
\end{align}
for $T>T_{BG}$.
Here, $\nu_1=\gamma_1/(4\pi\hbar^2v_0^2)$ is the density of states (DOS) per valley per
spin in bilayer graphene, $D_1\approx\unit[20-50]{eV}$ is the bare deformation
potential coupling, $D_2=\frac{\beta\hbar v_0}{2a}\approx\unit[7]{eV}$ the gauge potential
coupling \cite{Ochoa2011},
$s_{L}\approx\unit[2.1\times10^4]{m/s}$ the longitudinal
acoustic phonon velocity, and $\alpha =
e^2/(4\pi\epsilon_0\hbar{}v_0)\approx2$. Moreover, $(k_F\ell)_{MLG}=2\hbar^2v_0^2/(u^2n_0)$
is a dimensionless measure of short-range impurity concentration $n_0$ with $\delta$-function potential of
strength $u$.

Figure~\ref{fig:bi-J-comparison} summarizes the expected magnitudes of the different
contributions: the optical phonons are expected to dominate the heat flow at large temperatures.
Note that the bilayer electron--optical phonon coupling is around two orders of magnitude larger than in MLG near the Dirac point, when the renormalization of the Fermi velocity in suspended monolayer \cite{Elias2011}  is taken into account.


\begin{table}
\begin{tabular}{ c l c l c l c l c l c l c l c }
\hline{\smallskip}
S \#&$L$ &$W$ &$R_0$& $\sigma_m/\sigma_0$&$R_C $&$V_D$&$\mu_f$ \\
\hline {\smallskip}
S1 &  0.43  & 0.51  & 5.8 k$\Omega$ & 3.0 & 100 $\Omega$ &-0.3V&14000  \\
S2 &  0.83  & 3.2   & 1.5 k$\Omega$ & 3.6 & 30 $\Omega$   &0.6V&6200\\
\hline{\smallskip}
\end{tabular}
 \caption{Parameters for our two bilayer graphene samples denoted by S1 and S2. The length and width are given  in $\mu$m by $L$ and $W$, respectively. $R_0$ is the maximum resistance at the Dirac point which corresponds to minimum conductivity $\sigma_{m}$ as multiples of $\sigma_0$ = $\frac{4e^2}{\pi h}$, while $R_C$ is an estimate for the contact resistance. The last column indicates the field effect mobility (in cm$^2$/Vs) deduced from the gate sweeps. Sample S1 is an HF-underetched device on silicon dioxide whereas S2 was fabricated on LOR resist.}

\label{SAMPLEparams}
\end{table}

\begin{figure}
\centering
\includegraphics[width=0.9\linewidth]{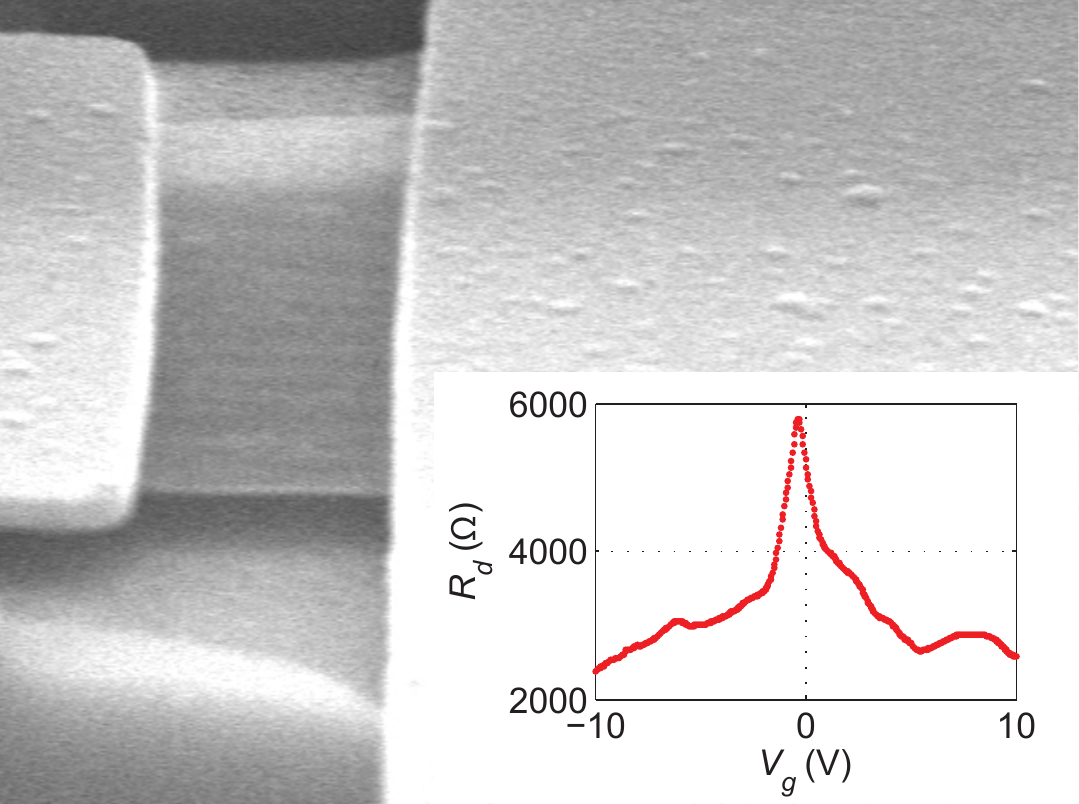}
\caption{Scanning electron micrograph of our suspended  graphene sample S2. The metallic leads for contacting graphene were made of Cr/Au. The inset displays zero-bias resistance versus $V_g$ for S1.  }
\label{fig:schema}
\end{figure}

The studied bilayer samples are listed in Table I:  their length varied over $L=0.4 - 0.8$ $\mu$m and the width $W=0.5 - 3.2$ $\mu$m. The samples studied here are exfoliated graphene suspended on SiO$_2$ (S1) and LOR resist (S2). Leads were patterned using e-beam lithography.
Raman spectroscopy was employed to verify that the sample was a bilayer graphene sheet. Prior to measurements, all samples were cleaned by current annealing using a current of $\sim 1$ mA/$\mu$m in cryogenic vacuum. The resulting high-quality samples were nearly neutral, with the charge neutrality (Dirac) point located at $|V_g^{D}| < 0.6$ V. The gate capacitance was determined from the parallel plate capacitor model: $C_g = 5.2-4.7 $ $\frac{nF}{cm^2}$ for different samples. A SEM image of sample S2 is shown in Fig.~\ref{fig:schema}. The graphene sheet has wavy structure, which indicates presence of tensile stress at room temperature. Strain may be the reason why no clear signs of flexural modes \cite{Ochoa2011}  were observed in the $I/V$ measurements on our bilayer samples (see Supplementary Material).

The inset in Fig. \ref{fig:schema} displays the variation of zero-bias resistance $R_0=dV/dI_{|V=0}$ vs. gate voltage $V_g$ over chemical potentials ranging $\pm 12$ meV across the Dirac point. The gate sweep  indicates that minimum charge density is around $1 \cdot 10^{10}$ cm$^{-2}$ in our bilayer samples.
The initial slope of $G_0 (n)=1/R_0$ was employed to determine the field effect mobility which reached $\mu_f =  1.4 \cdot 10^4$ cm$^2$/Vs in sample S1.

\begin{figure}
\centering
\includegraphics[width=0.9\linewidth]{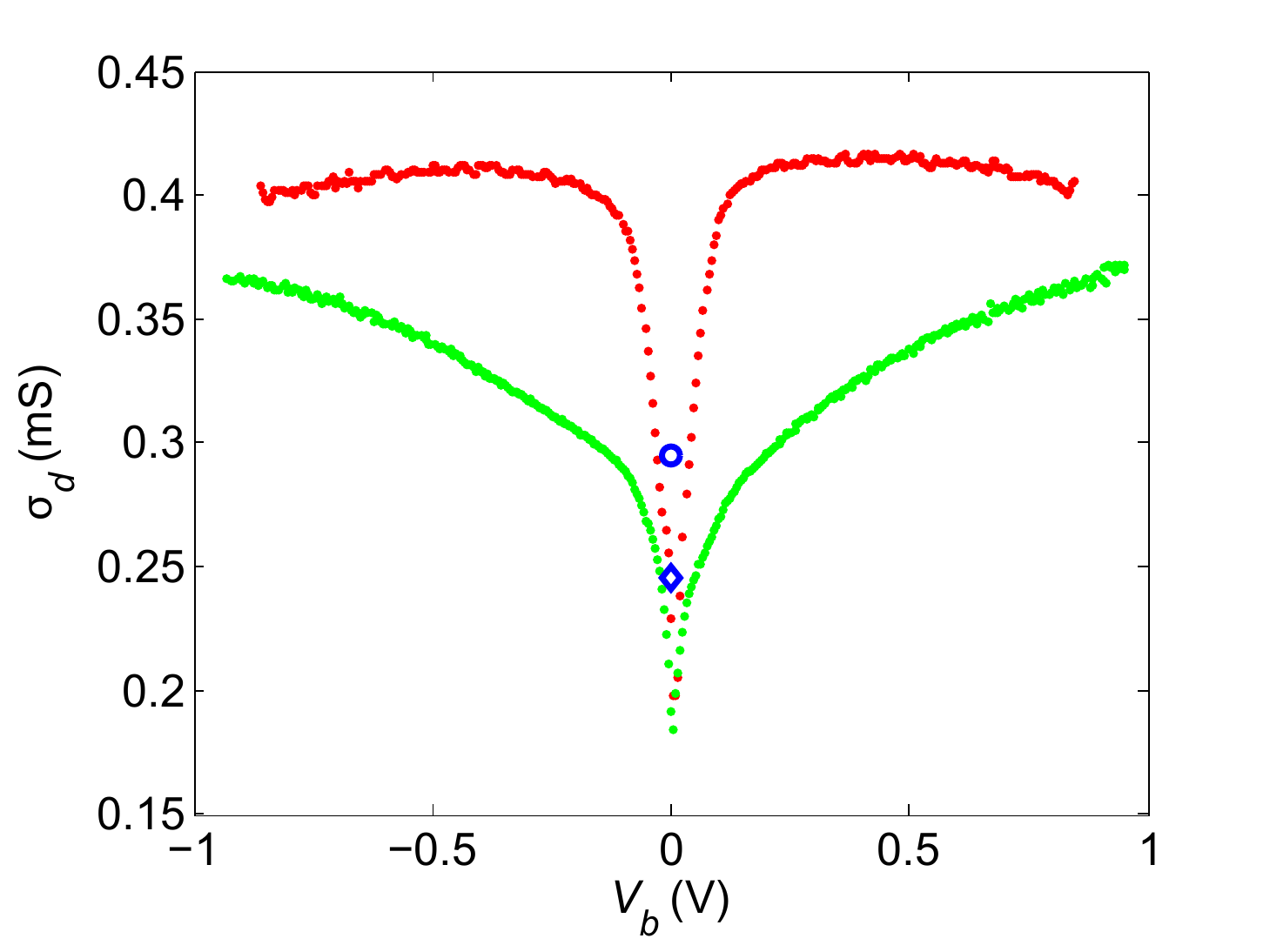}
\caption{a) Differential conductivity $\sigma_d$ vs. bias voltage $V_b$ near the Dirac point at $V_g=+0$ V for S1 (red, upper) and S2 (green, lower). The theoretical result for bilayer $24\frac{e^2}{\pi h}$ \cite{Cserti2007} is marked by $\circ$; the experimental result by Mayorov et al. \cite{mayorov2011} on suspended bilayer is denoted by $\lozenge$.
}
\label{fig:results}
\end{figure}

Fig. \ref{fig:results}  displays differential conductance \textit{vs.} bias voltage for the samples. Initially, there is a strong increase in conductance \textit{w.r.t} bias voltage, which has been interpreted as a self heating of the bilayer \cite{Viljas2011}. The increase is cut off at bias values close to optical phonon energy \cite{Fay2011}. On the other hand, flexural phonons were found to play a role in a monolayer by facilitating supercollisions at high bias; their presence was observed in the total resistance as a temperature dependent contribution which behaved as $V/I \propto T^2$ \cite{Laitinen2014}. In the Supplementary material we show that the integrated data of Fig. \ref{fig:results} yields $V/I \propto T_e$ and no indications of flexural modes are observed. The most likely explanation for this is that strain changes the dispersion relation of flexural modes from quadratic to linear, making the $T^2$ component negligible.  The measured minimum conductivity is around $\frac{13e^2}{\pi h}$ and quite close to the results of Mayorov et al. \cite{mayorov2011} on similar suspended bilayer samples.

Besides current voltage characteristics, the zero frequency shot noise (600-900MHz) and its first harmonic was measured to deduce the electronic temperature of BLG under high bias.  The basic argument for shot noise thermometry are found in Refs.~ \onlinecite{wu2010,santavicca2010} and its recent applications in graphene are found in Refs.~\onlinecite{Betz2012, Laitinen2014}. Noise spectral density of a Poissonian process is given by $S_p = 2q\langle I\rangle$, where $\langle I\rangle$ is the average current. The Fano factor defines the noise level $S_I$ with respect to the Poissonian noise, $F = S_I /S_p$ \cite{Blanter2000}. In addition to gate voltage, the Fano factor depends on the bias voltage $V$ (see the Supplementary Material). When the bias voltage is increased, enhanced electron-electron interactions try to bring the system towards the so called "hot electron regime" with $F=\sqrt{3}/4$ whereas inelastic scattering causes tendency towards classical behavior without any shot noise.

Shot noise thermometry was used to determine the temperature of electrons in graphene under Joule heating. The temperature, $T_e = \frac{Fe|V|}{2k_B}$, is an average over spatial distribution of temperature which, in the regime of strong electron-phonon scattering, yields rather accurately the actual electronic temperature in the center of the sample; the Fano factor for noise thermometry was adjusted for the contact resistance as in Ref. \onlinecite{Laitinen2014}. Our experiments were performed on a pulse-tube-based dilution refrigerator operated around 0.5 K. For experimental details, we refer to Refs.~\onlinecite{danneau2008b,Oksanen2014}.

\begin{figure}
\centering
\includegraphics[width=1.0\linewidth]{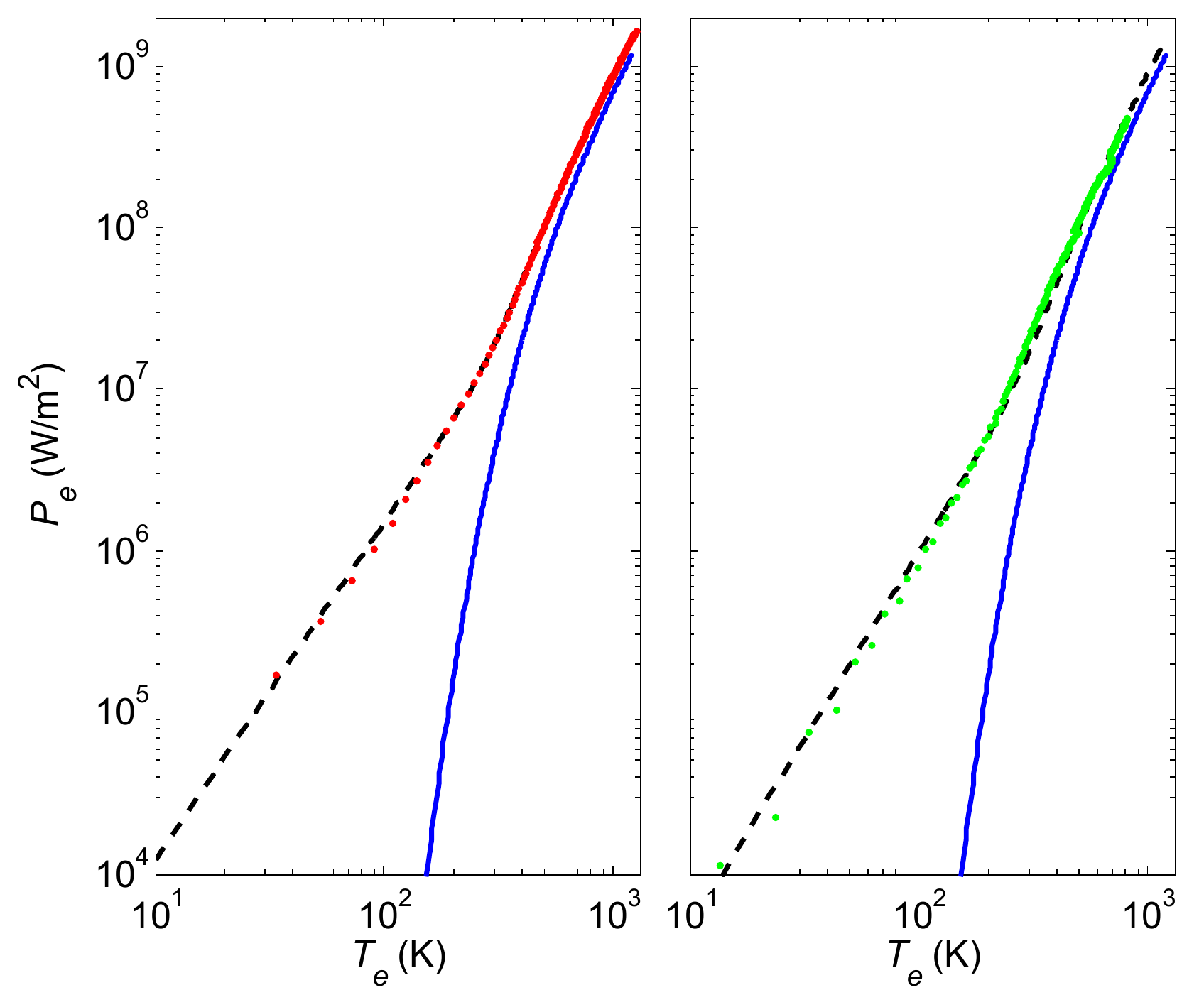}
\caption{Joule heating in bilayer graphene $P_e$ as a function electron temperature $T_e=Fe|V|/2k_B$ near the Dirac point at $n$ = $10^{10}$ $\frac{1}{cm^2}$. The  left (sample S1) and the right (sample S2) frames represent measured data in circles while the theoretical behavior is expressed using dashed black curves; the blue curves denote the contributions of optical phonons.}
\label{fig:combined}
\end{figure}

The Joule heating power to graphene electrons equals to $P_e = VI- R_C I^2$ where $R_C$ denotes the effective contact resistance (see Table I). Fig. \ref{fig:combined} displays $P_e$ vs. $T_e$ measured up to $10^9$ W/m$^2$.
Due to weak e-ph coupling at low bias, the electronic heat conduction governed by the Wiedemann-Franz law dominates other thermal processes. The power $ P_{WF}= \frac{\pi^2}{6}\frac{k_B^2}{e^2} \left( T_e^2-T_B^2 \right)$ is carried to the leads at temperature $T_B$. On the other hand, at large bias, e-ph coupling is strong and hence heat transport to phonons ($P_{e-ph}$) dominates. Due to the short length of the samples investigated here, $P_{e-ph}$  is seen to dominate the measured $P_e(T_e)$ first above $200-300$~K . For acoustic phonons temperature dependence of $\propto T^2$ is expected in bilayer ($\delta =2$), but the theoretical estimates (see Fig. \ref{fig:bi-J-comparison}) indicate power flows that are well below the observed levels unless the deformation potential is made exceedingly large. Above 300 K, the results indicate an onset of additional relaxation channel which initially leads to steeper $T$ dependence, but whose growth rate becomes slightly weaker with growing temperature. This onset behavior is similar to that calculated for optical phonons in Fig. \ref{fig:bi-J-comparison} (see also the fits in Fig. \ref{fig:combined}). As the data sets coincide at high temperatures, the coupling we measure is truly an intrinsic property of bilayer graphene; in the Wiedemann-Franz regime at low temperatures, the behavior at different lengths deviate from each other because of scaling as $1/L^2$.

In the high-$T$ regime where the phonon scattering dominates the electronic heat diffusion, we find quite weak dependence of $P_{e}$ on chemical potential (see Fig. S6 of the Supplementary Material). In the measured range (-12 meV $< \mu < $ 12 meV), all the variation of the absorbed heat flux $P_e$ at constant $T_e$ can be accounted for by a change in $P_{WF}$ due to variation of $R(V_g)$. The independence of electron--phonon coupling on $\mu$ is in agreement with optical or acoustic phonons in BLG with $T_{ac} > T_{BG}$ and $k_B T_e >> \mu $. The observed temperature dependence of the coupling, however, rules out the latter possibility as in that regime $ {P}_{e-op}\propto T_e^2-T_{ph}^2$ \cite{Viljas2010}. We note that the weak gate dependence of $P_{e-ph}$ does not rule out supercollisions as they predict  $\mu$ independent behavior due to constant density of states in bilayer. The best check for supercollision scattering is to scrutinize $P_e/T_e^3$, which shows no plateau and so rules out the supercollision processes \cite{Betz2012} (see Fig. S5 in the Supplementary material).

Considering only optical phonons, the data in Fig.~\ref{fig:combined} is compared to the sum
$P_{WF}+P_{e-op}^{(ZC)}+P_{e-op}^{(ZE)}$,
using parameters specified below Eq.~\eqref{e.blqlto}. $\gamma_0'$ is left as a fit parameter. In our experiments at weak
doping, the best fit is obtained with $\gamma_0'\approx 37$ eV/nm. This is slightly smaller than the value $\gamma_0'=42$~eV/nm obtained in Ref.~\onlinecite{Yan2007} and $\gamma_0'=47.5$~eV/nm found in Ref. \onlinecite{Mauri2014}. Using $\gamma_0'=42$~eV/nm, our results would imply nearly equal contributions from ZC and ZE phonons to heat current. Altogether, the optical phonons together with the Wiedemann-Franz law give a very good agreement between experiment and theory as indicated by the fits in Fig. \ref{fig:combined} (see also Fig. S5 in the Supplementary Material).

%


Compared with monolayer experiments \cite{Betz2012a,Laitinen2014}, our results on the electron-phonon coupling are rather close in magnitude in the degenerate limit ($\mu > k_B T$) at high bias. At small chemical potentials, the constant density of states in bilayer implies stronger phonon scattering in bilayer. According to Ref. \onlinecite{Viljas2010}, optical phonon heat flow is larger by a factor of three in bilayer than in monolayer, and this difference  grows further with Fermi level renormalization near the Dirac point. Furthermore, the rigidity of the bilayer is larger than for monolayer \cite{Ochoa2011} which, on its part,  diminishes the strength of the flexural phonon supercollisions in bilayer. These issues account for the fact that only in bilayers we are able to observe intrinsic scattering by optical phonons, while flexural-mode-induced supercollisions appear in suspended monolayers.

In summary, our experiments indicate strong difference between electron-phonon heat relaxation at high bias in suspended monolayer and bilayer graphene. Using electrical transport experiments and shot noise thermometry, we find that electron-optical phonon scattering dominates in bilayer graphene at electronic temperatures of $300-1000$ K, induced by bias voltages comparable to optical phonon energies. The strength of the scattering follows theoretical expectations with a specific thermal activation behavior, and indicates the presence of intervalley electron scattering by zone edge and zone center optical phonons. This electron-phonon coupling is found to be independent of the gate-induced chemical potential at $|\mu| < 12$ meV, which is in accordance with the theory for optical phonon scattering.

We acknowledge fruitful discussions with J. Viljas, T. Heikkil\"{a}, and F. Mauri.  Our work was supported by the Academy of Finland (contracts no. 135908 and 250280, LTQ CoE). The research leading to these results has received funding from the European Union Seventh Framework Programme under grant agreement no 604391 Graphene Flagship, and the work benefitted from the use of the Aalto University Low Temperature Laboratory infrastructure. MO is grateful to V\"{a}is\"{a}l\"{a} Foundation of the Finnish Academy of Science and Letters for a scholarship.

\bibliography{Collection}

\section{\large{Supplement}}
\section{Sample fabrication and characterization}

\label{sec:sampleFabrication}
Initially, we employed HF-etching in order to make graphene suspended  \cite{Bolotin2008}, but later we switched to the technique introduced in Ref. \onlinecite{Tombros2011}, the LOR-technique. This technique was selected over the conventional HF-technique, because of the lower parasitic pad capacitance that can be achieved, which is important for our microwave measurements. The lower parasitic capacitance is due to longer distance between the graphene circuit and the strongly doped $Si^{++}$ back gate.

\begin{figure}[htp]
\centering
\includegraphics[width=0.9\linewidth]{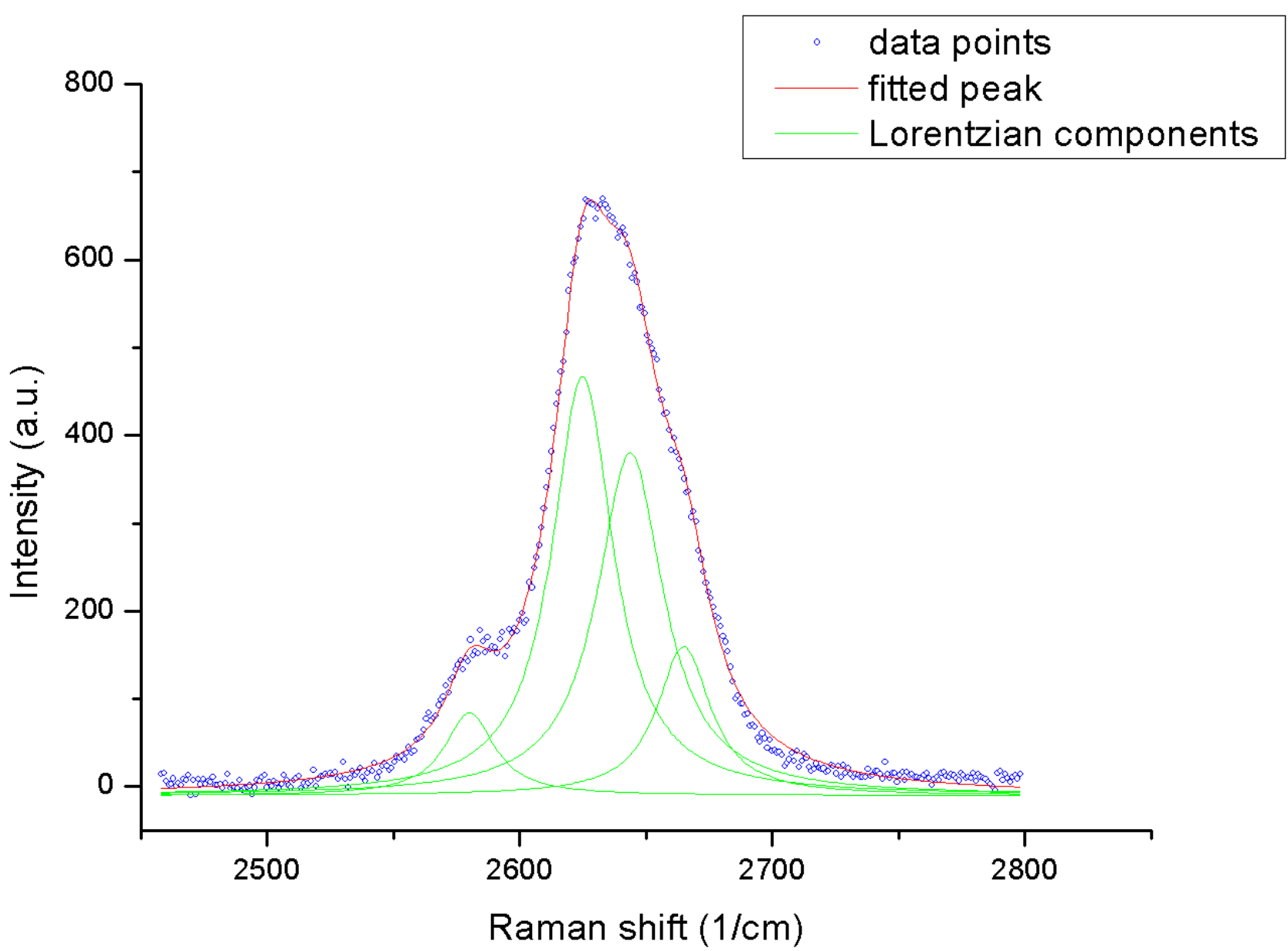}
\caption{2D peak in a Raman spectrum of a suspended sample fabricated with the LOR-method. The red trace displays a sum of four fitted Lorentzian indicated in green.
}
\label{fig:RamanSpectrum}
\end{figure}

Compared to a Raman spectrum of a sample directly on $SiO_2$ with the standard D, G and 2D peaks, there are some additional features in a spectrum of a sample on LOR. Near 1600 1/cm, where one expects to see G-peak characteristic for carbon and 1350 1/cm D-peak associated with defects, there are a few other peaks. Origin of these peaks is unclear, but since the sample performance with these LOR-samples has been good compared to HF-etched samples, it can be assumed that the peaks originate from the LOR-layer. Additionally, the D-peak is virtually nonexistent which implies very few defects in the graphene, as verified by the high mobility of the samples. Around 2600 1/cm, where one observes the 2D peak used to determine the number of layers, there are no excess peaks visible. However, the measured widths of the 2D-peaks for samples on LOR tend to be slightly larger than expected (\emph{i.e. }in comparison with the samples on $SiO_2$). Fig. S1 displays the 2D peak of a bilayer sample suspended using the LOR technique.

\section{Noise}
\begin{figure}[htb]
\centering
\includegraphics[width=0.8\linewidth]{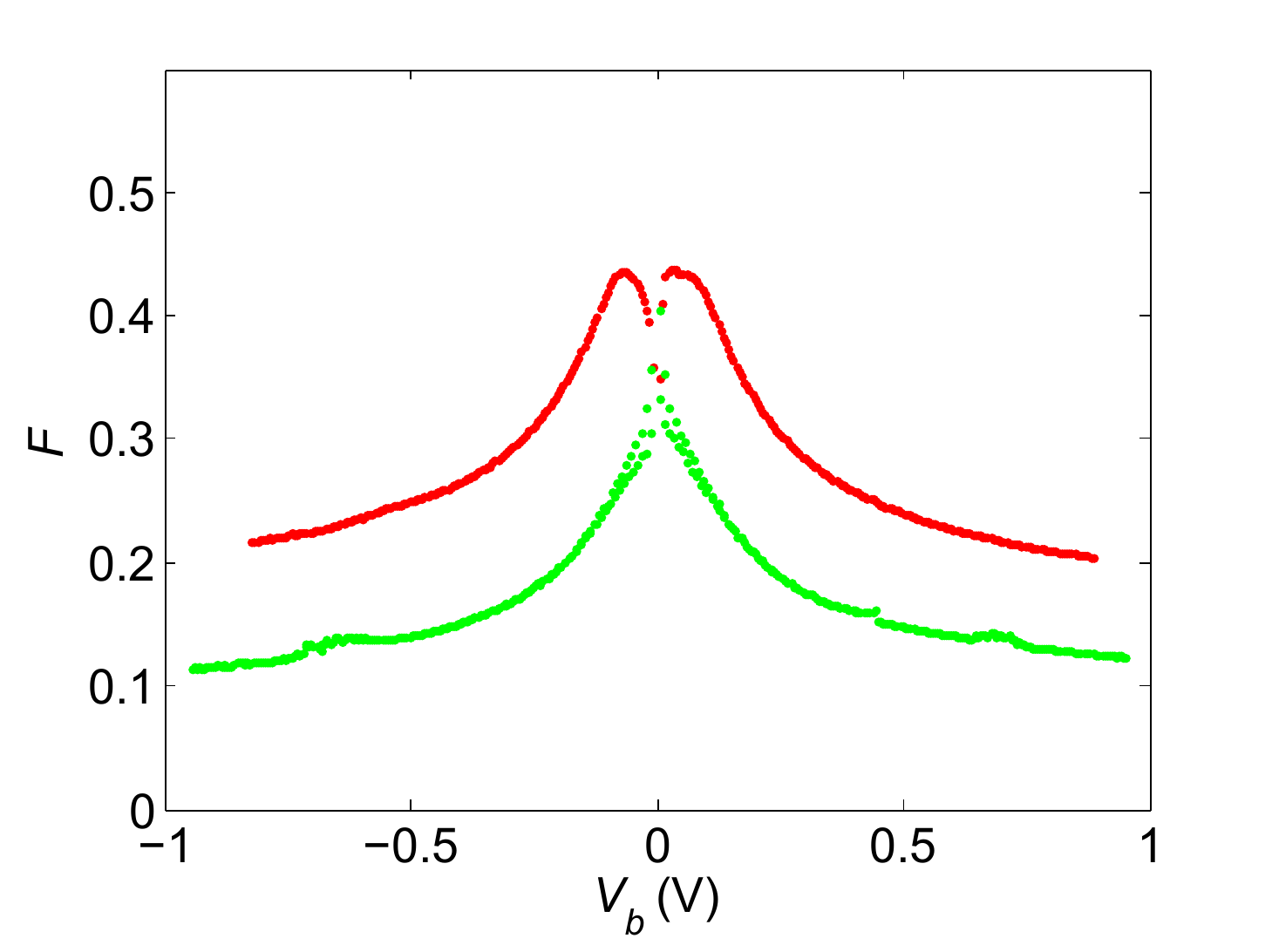}
\caption{Excess Fano factor $F$ vs. bias voltage $V$ near the charge neutrality point at charge density $n \simeq 1.0 \cdot 10^{10}$ cm$^{-2}$. The red curve represents sample S1 and the green one corresponds to sample S2.}
\label{fig:noise}
\end{figure}

  Shot noise $S_I$ measured up to high bias is illustrated in terms of the Fano factor in Fig. \ref{fig:noise}. The Fano factor is determined for excess noise: $F=\left(S_I (I)-S_I (0)\right)/2eI$, where $S_I (I)$ and $S_I (0)$ denote the current noise power spectrum at current $I$ and at $I=0$, respectively.  Near the Dirac point, there is a slight initial increase of $F$ in sample S1, which reflects tendency towards the hot electron regime as the effective strength of the electron-electron interaction becomes enhanced with growing bias \cite{steinbach1996,nagaev1995,Sarkar2014}. On the other hand, the Fano factor of sample S2 indicates an immediate decrease with bias, which is a sign of inelastic processes entering already at low energies, for example due to sliding modes in bilayer graphene \cite{Louie2008}; this inelastic scattering is observed even in the differential conductance of sample S2 in Fig. 3 of the main paper where the linear increase with bias voltage is cut off earlier than in sample S1. Pretty smooth, symmetric decrease in $F$ is observed with the bias voltage over the full range of bias conditions; moreover, no difference in $F$ at equal hole and electron density was found so that there is no difference in e-ph coupling for the electron and hole transport regimes.

\section{Details of the electrical characteristics}
\begin{figure}[htb]
\centering
\includegraphics[width=0.8\linewidth]{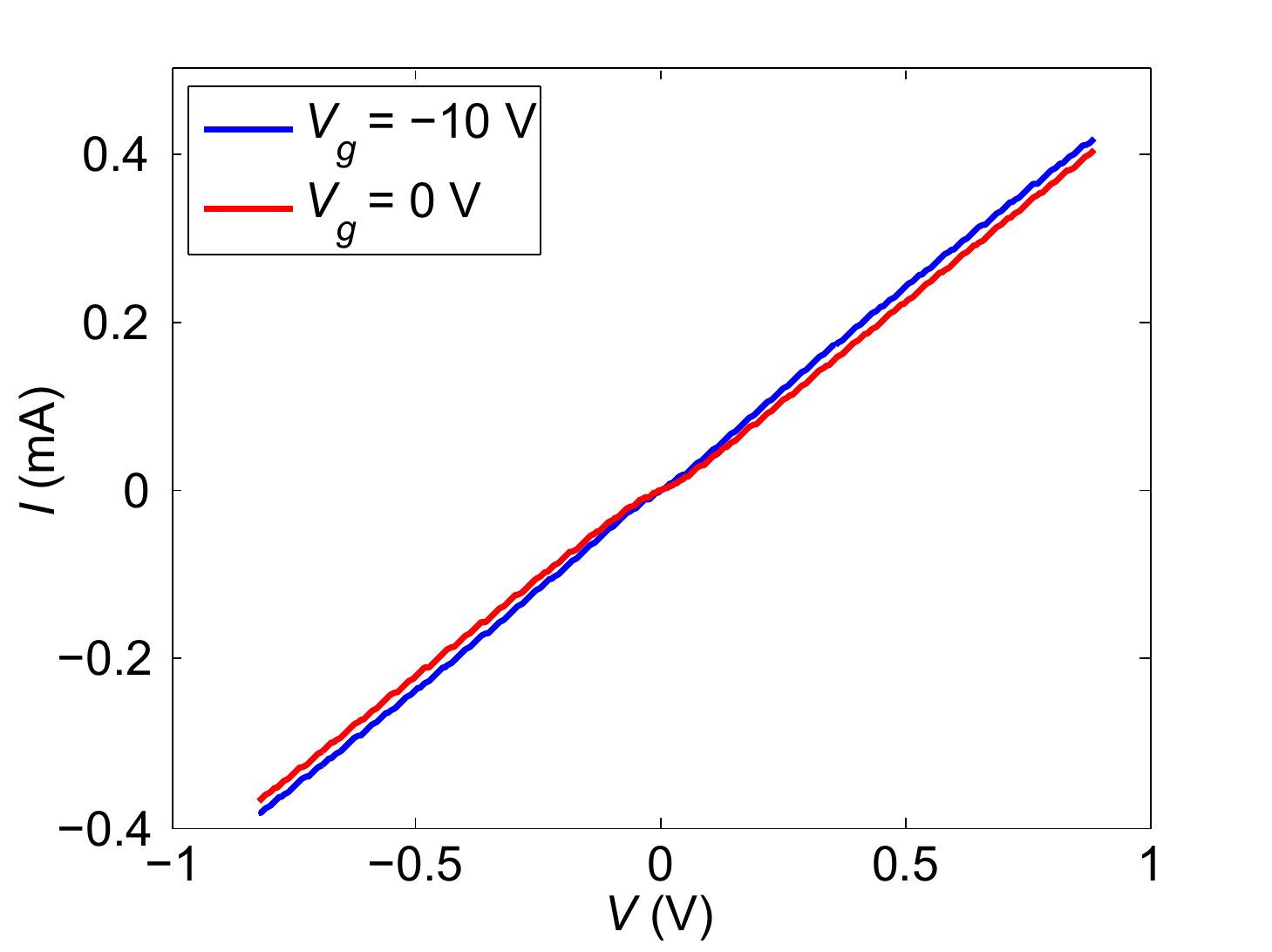}
\caption{Current $I$ vs. bias voltage $V$ around the Dirac point at charge density $n = 1.0 \cdot 10^{10}$ cm$^{-2}$ (red) and $n = 2.8 \cdot 10^{11}$ cm$^{-2}$ (blue) for sample S1.}
\label{fig:IV}
\end{figure}

\begin{figure}[b]
\centering
\includegraphics[width=0.8\linewidth]{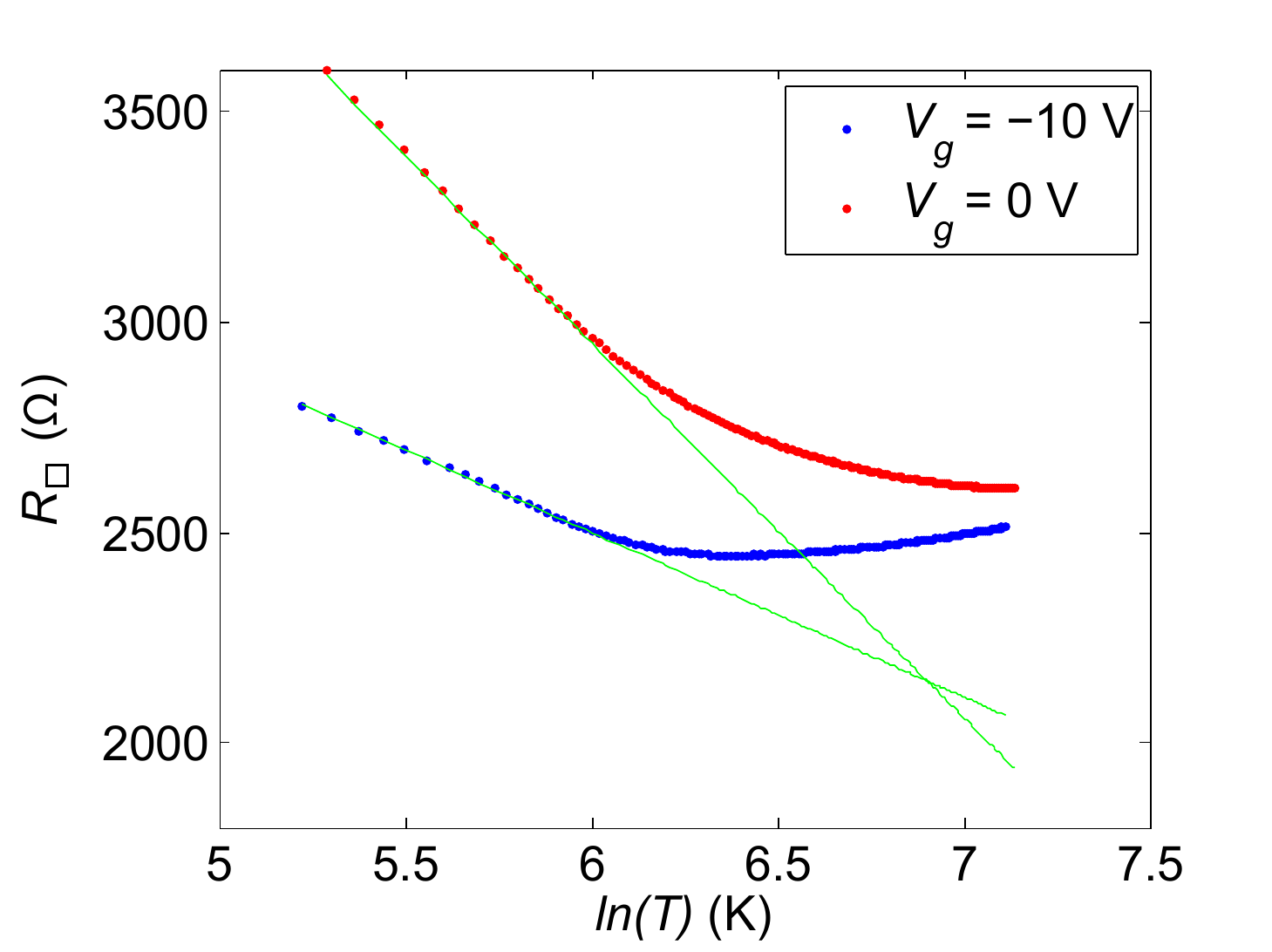}
\includegraphics[width=0.8\linewidth]{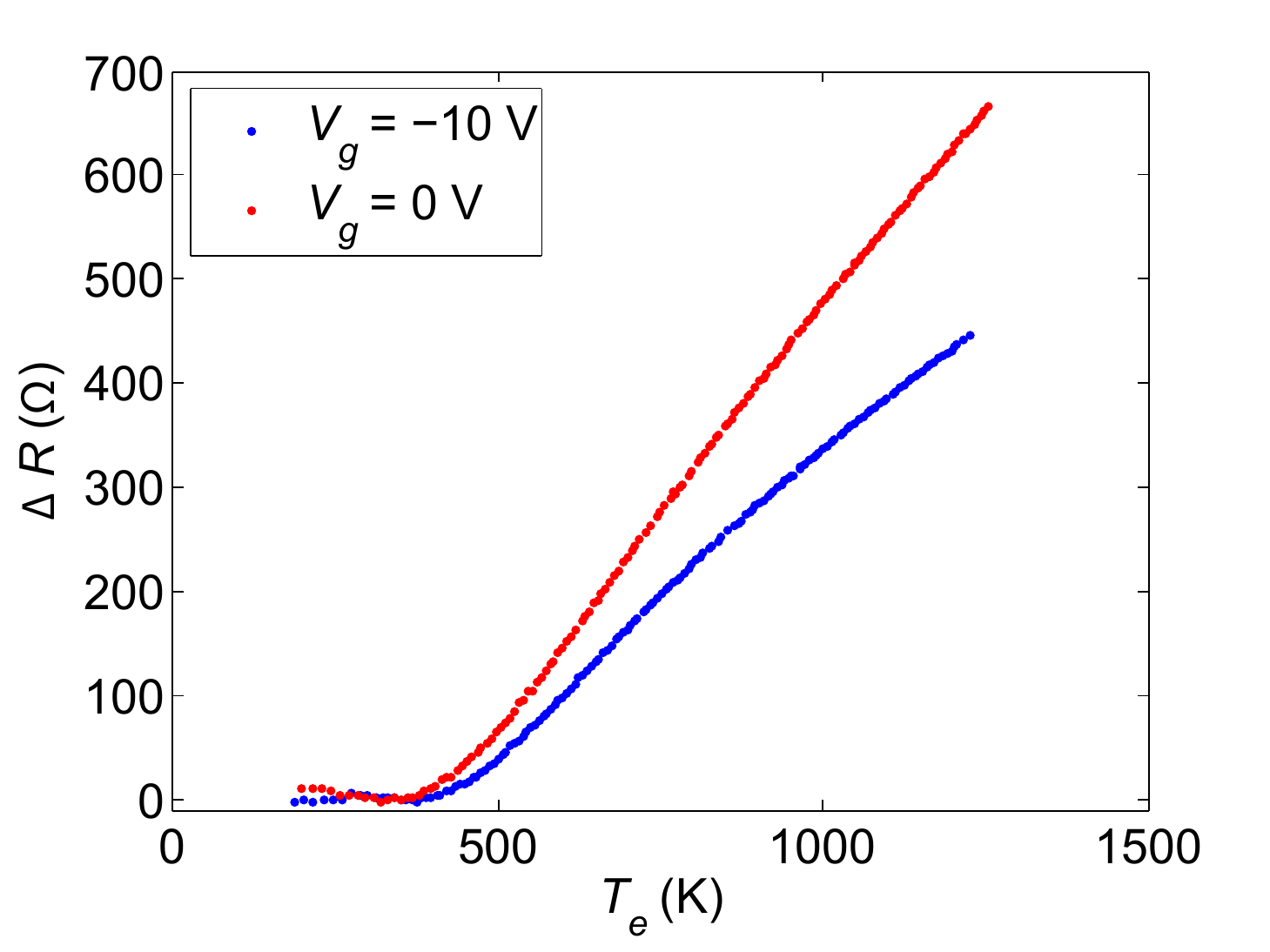}
\caption{a) Total sheet resistance $R_{\square}=(V/I) W/L$ as a function of the logarithm of the electronic temperature deduced from the Fano factor using $T_e=Fe|V|/2k_B$ (at $1.0 \cdot 10^{10}$ cm$^{-2}$ and at $|n| = 2.8 \cdot 10^{11}$ cm$^{-2}$) for sample S1. The green lines indicate logarithmic temperature dependence at low temperature. b) Deviation from the logarithmic behavior  $R_{\square}-C \log(T_e)$ for the two data sets in the upper frame.}
\label{fig:RvsT}
\end{figure}

In our experiments, we measured the electrical characteristics of the samples both at DC and at low-frequency AC (dynamic resistance $R_d= dV/dI$).
IV-curves of S1 measured at the charge density $n = 1.0 \cdot 10^{10}$ cm$^{-2}$ and $n = 2.8 \cdot 10^{11}$ cm$^{-2}$ are illustrated in Fig. \ref{fig:IV}. The IV-curve displays a clear decrease in $R_{\square}=(V/I) W/L$ with growing bias voltage $V$ around the Dirac point. The differential resistance $R_d= dV/dI$, measured by lock-in methods, corresponds to the inverse slope of the IV-curve. The $R_d(V )$ measurements were employed to determine the coupling strength of the current noise from the sample to the preamplifier at microwave frequencies.
The behavior of $R_{\square}$ as a function of shot noise temperature $T_e$ is illustrated in  Fig.  \ref{fig:RvsT}  at the same charge densities as the IV curves in Fig. \ref{fig:IV}. The temperature dependence of $R_{\square}$ is well fit at $T_e \lesssim 100$ K using $C\log(T_e)$ where the prefactor $C \simeq 390 - 890$ $\Omega$. When the fitted $C\log(T_e)$ part is subtracted off from $R_{\square}$, we obtain for the difference $\Delta R_{\square}$ an almost linear dependence  on $T_e$. This behavior suggests that the scattering rate by optical phonons grows nearly linearly above $300-400$ K, which is in agreement with recent single layer calculations of Ref.  \onlinecite{Mauri2014} without electron-electron interaction effects and with the non-suspended bilayer analysis of Ref. \onlinecite{Fay2011}.

Note that linear $R(T)$ is predicted for in-plane acoustic phonons (see e.g. Ref. \onlinecite{Hwang2008}) but these cannot be the dominant processes because of the results on the electron phonon coupling. The $\log(T_e)$ dependence may be due to very robust weak localization effects \cite{Savchenko2008} or it may be  a signature of increased effective disorder\cite{Beenakker2008scatt}; the disorder may cause logarithmic increase in conductance as a function of (disorder length scale)$^{-1}$ .

\section{Analysis of the power laws}
\begin{figure}[hbt]
\centering
\includegraphics[width=0.8\linewidth]{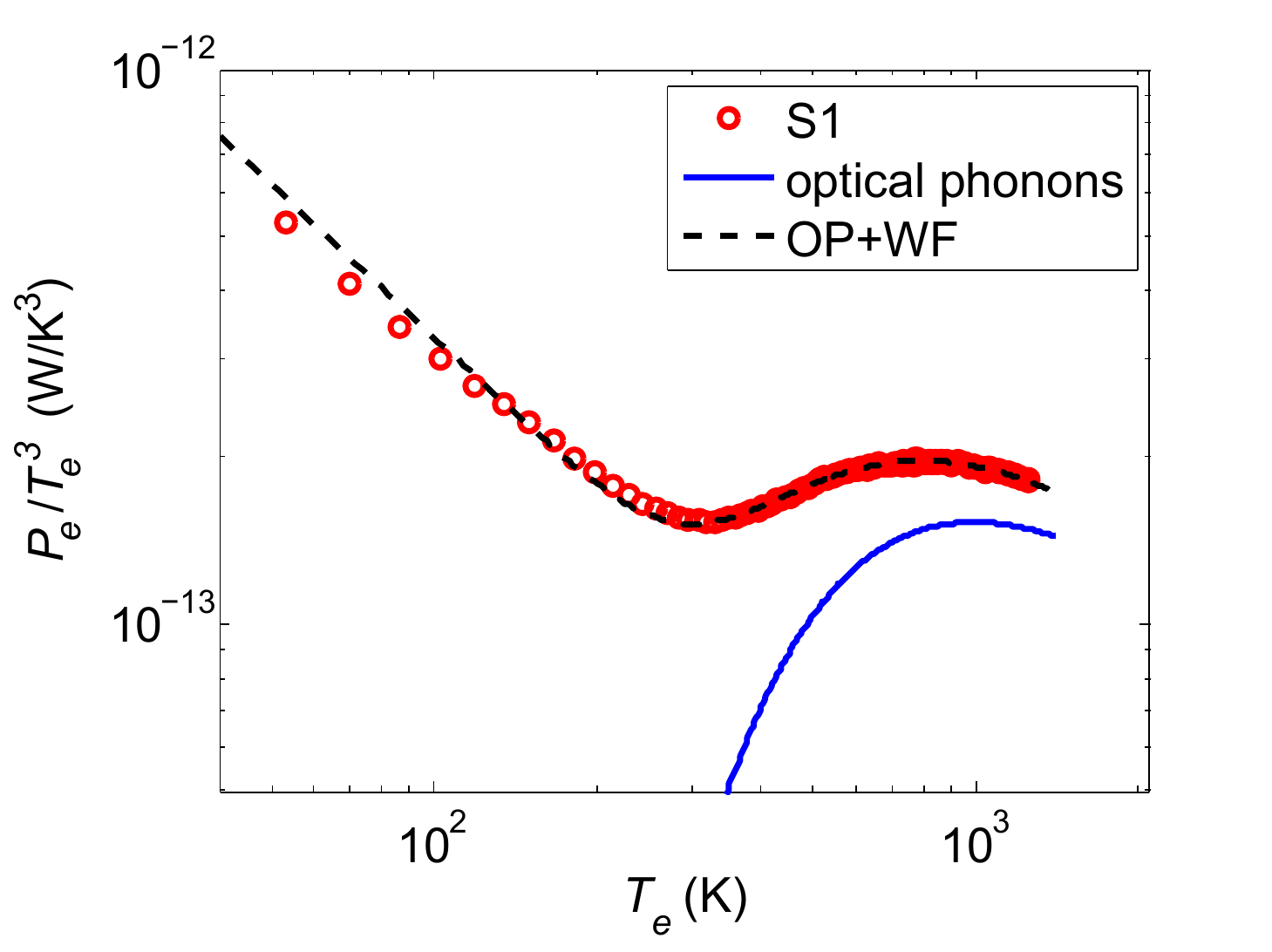}
\caption{Measured heat flow from electrons to phonons (data S1 from Fig. 4 in the main text) normalized by $T_e^3$ and displayed as a function of $T_e$. The dashed curve illustrates the theoretical curve from Fig. 4 (left frame) of the main paper. The blue curve denotes the heat flow due to the optical phonon scattering alone, while the dashed curve contains additionally the electronic heat conduction.}
\label{fig:PvsT3}
\end{figure}

In order to support the conclusion of electron - optical phonon scattering, let us plot the data of Fig. 4 in the main paper in a slightly different form. For acoustic phonons or supercollision processes, the heat flow from electrons to the phonons  would be approximately characterized by the power law $P \propto T_e^{\delta}$ (here we have dropped the small  phonon temperature term $T_{ph}^{\delta}$) where $\delta$ would be 2 or 3, respectively. The law with $\delta = 2$ can  be ruled out immediately without any doubt. For the supercollision case  with $\delta = 3$,  we compare our results with the theory by replotting our data in the same way as in the experimental works on supercollisions in graphene \cite{Betz2012, Betz2012a,Laitinen2014}:  Fig. \ref{fig:PvsT3} displays the original data of Fig. 4 in  the main paper by normalizing the power flow $P_e$ with $T_e^3$. No clear saturation is observed to  signify  $\delta = 3$ behavior. As a reference we have plotted the behavior for electron - optical phonon scattering, which reproduces the asymptotic features of the data. Clearly, the shape of the curve follows the thermally activated optical phonon tendency which is characteristic to the formation of modes with well defined energy.

\begin{figure}[H]
\centering
\includegraphics[width=0.8\linewidth]{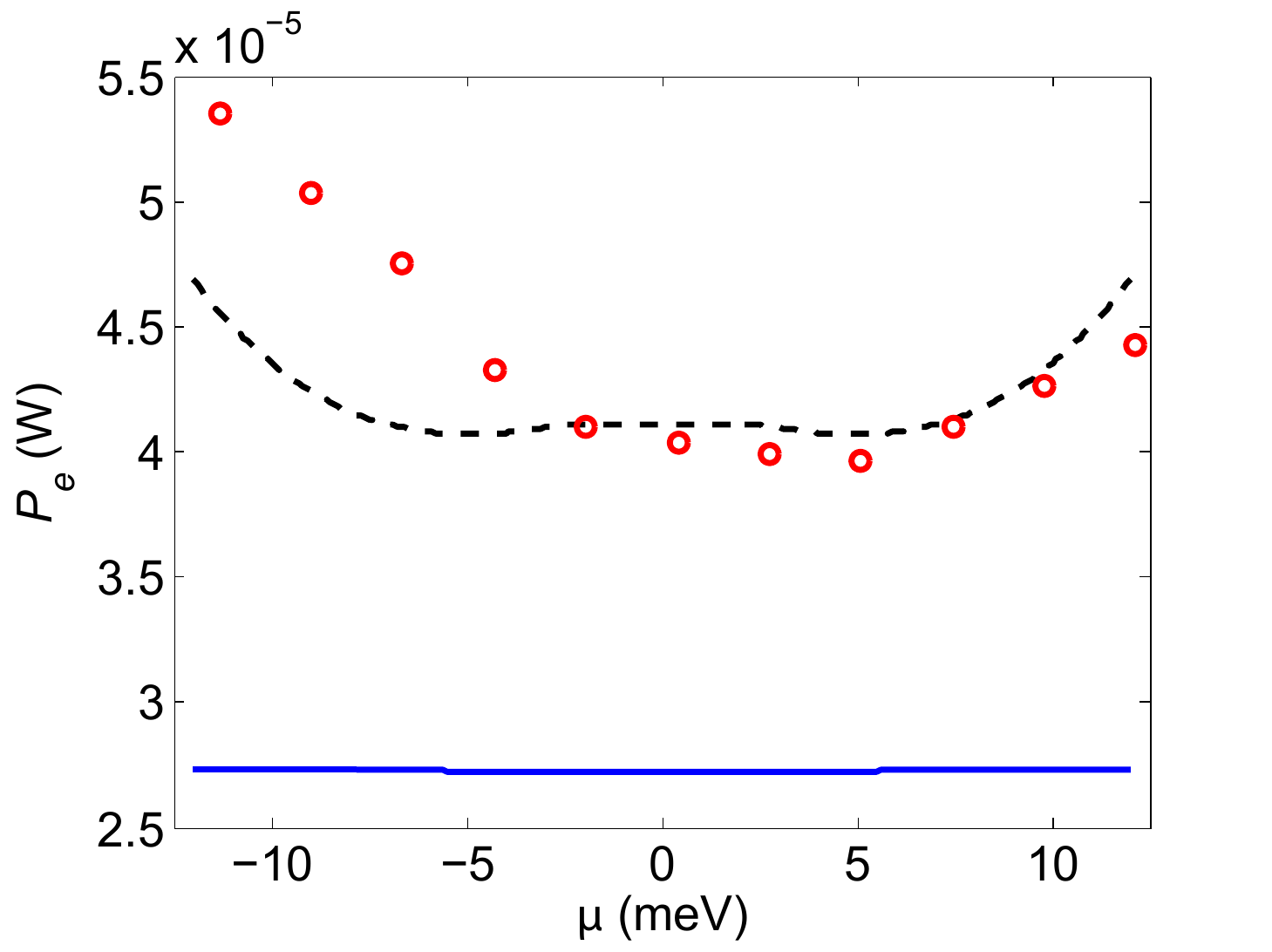}
\caption{Chemical potential dependence of the measured heat flow from electrons to phonons at $T_e=600$ K  for sample S1. The dashed line indicates the theoretical variation obtained using the change in the electronic heat conduction on top of the optical-phonon-facilitated heat flow which is denoted by the blue curve \cite{Bistritzer2009,Viljas2010}. }
\label{fig:mudep}
\end{figure}

The dependence of the electron phonon coupling on the chemical potential is illustrated in Fig. \ref{fig:mudep} for sample S1. We find a weak variation with chemical potential which we assign to the variation in electrical resistance of the sample. The dashed curve in Fig. \ref{fig:mudep} displays the sum of the theoretical electron phonon coupling and the electronic part for heat conduction which was obtained from the measured electrical resistance  $R(V_g)$, symmetrized at high bias, and the Wiedemann-Franz law. The slightly stronger measured variation compared with the theoretical curve might be an indication of mass renormalization due to interactions \cite{Borghi2009} although these effects are expected to be small.

\bibliography{Collection}

\end{document}